\documentstyle[twocolumn,prl,aps,epsf]{revtex}

\begin{document}

\twocolumn[\hsize\textwidth\columnwidth\hsize\csname
@twocolumnfalse\endcsname

\title{Persistent currents in mesoscopic rings with a quantum dot}
\author{A. A. Aligia}
\address{Comisi\'{o}n Nacional de Energ\'{\i}a At\'{o}mica,
Centro At\'{o}mico Bariloche and Instituto Balseiro,
8400 Bariloche, Argentina}
\maketitle

\begin{abstract}
Using the Anderson model in the Kondo regime, we calculate the persistent
current $j$ in a ring with an embedded quantum dot (QD) as a function of the
Aharonov-Bohm flux $\Phi $ for different ring length $L$, temperature $T$
and broadening of the conduction states $\delta $. For $T=\delta =0$ and 
$L\gg \xi$, where $\xi $ is the Kondo screening length, $Lj$ tends to the
value for a non interacting ideal ring, while it is suppressed for a side
coupled QD. For any $L/\xi$, $Lj$ is also suppressed when either $T$ or 
$\delta $ increase above a fraction of the level spacing which depends on 
$\Phi $.
\end{abstract}


] 

\narrowtext

Electron transport through a quantum dot (QD) has been a subject of great
interest in the past few years. The progress in nanofabrication made it
possible to use QD's as ideal realizations of the Kondo effect, which is one
of the most exciting and studied problems in condensed matter physics. It
consists in the screening of an impurity spin by a cloud of conduction
electrons of radius $\xi \sim \hbar v_{F}/T_{K}$ and energies $\sim 2T_{K}$
around the Fermi energy $\varepsilon _{F}$, where $T_{K}$ is the Kondo
temperature and $v_{F}$ is the Fermi velocity. The energy scale $T_{K}$ is
experimentally accessible in different ways, like the width of the peak in
linear response conductance through a QD and its temperature dependence \cite
{wiel}. Instead, a direct measurement of $\xi $ does not exist so far.

Several interesting experiments were performed recently in an Aharonov-Bohm
geometry, in which a ring containing a QD is threaded by a magnetic flux 
\cite{wiel,schu,hei}. Phase coherence along the ring has been demonstrated.
The persistent current $j$ in these rings, and in rings side-coupled to a QD
has been studied theoretically \cite{zv1,fer,kang,aff,hu,eck,anda,zv2}. 
However, basic
results of these works contradict each other and an accurate method to
calculate $j$ for any $L/\xi $ has not been developed. 
Exact Bethe ansatz results were known for $L\gg \xi $ and chiral electrons 
\cite{zv1}. These were extended to electrons moving in both directions 
\cite{eck}. The precise geometry of these calculations was explained
recently \cite{zv2}.
Perturbative renormalization group (RG) calculations have established the
form of $j(\Phi )$ for $L\ll \xi $ and $L\rightarrow \infty $ \cite{aff}.
A change in the
dependence of $j$ with magnetic flux is expected between the regimes $L\ll
\xi $ and $L\gg \xi $ \cite{kang,aff,hu}. Thus, measurements of $j$ would
provide a way of detecting $\xi $ \cite{aff}. Since the average level
spacing is $D=2\pi \hbar v_{F}/L$, the condition $L\sim \xi $ is equivalent
to $D/2\pi \sim T_{K}$.

Our purpose is to describe $j(\Phi )$ accurately through the crossover
region, and to consider the effects of temperature $T$ and finite level
width $\delta $ of the conduction states for the first time. 
We relate $j$ with the
one-particle Green function at the dot and the latter is calculated using an
interpolative perturbative approach (IPA) \cite{levy,kaj,pro}. For small $L$
and $T=\delta =0$, we also calculate $j$ using numerical exact
diagonalization (ED) finding excellent agreement with the IPA.

The Hamiltonian for the embedded dot is:

\begin{eqnarray}
H &=&-\sum_{\sigma ,j=0}^{L-1}\left( t_{j}e^{i\varphi _{j}}c_{j+1\sigma
}^{\dagger }c_{j\sigma }+\text{H.c.}\right)  \nonumber \\
&&+E_{d}\sum_{\sigma }n_{d\sigma }+Un_{d\uparrow }n_{d\downarrow },
\label{he}
\end{eqnarray}
with $c_{L\sigma }=c_{0\sigma }$ and $n_{d\sigma }=c_{0\sigma }^{\dagger
}c_{0\sigma }$. The phases $\varphi _{j}$ depend on the choice of gauge and
satisfy $\sum_{j}\varphi _{j}=\Phi $, where $\Phi hc/2\pi e$ is the magnetic
flux threading the ring. The hoppings are all equal except those involving
the QD: $t_{j}=t_{R}$ (right) for $j=0$, $t_{j}=t_{L}$ (left) for $j=L-1$,
and $t_{j}=t$ otherwise. The total current flowing between sites $l$ and $%
l+1 $ is:

\begin{equation}
j_{l}(\Phi )=\frac{e}{\hbar }\frac{\partial \langle H\rangle }{\partial
\varphi _{l}}=\frac{ie}{\hbar }t_{l}\sum_{\sigma }\langle e^{i\varphi
_{l}}c_{l+1\sigma }^{\dagger }c_{l\sigma }-\text{H.c.}\rangle .  \label{j}
\end{equation}
The expectation value entering Eq. (\ref{j}) is given in terms of Green
functions $\langle \langle c_{i\sigma };c_{j\sigma }^{\dagger }\rangle
\rangle .$ These in turn can be expressed in terms of the QD Green function $%
G_{d\sigma }(\omega )=\langle \langle c_{0\sigma };c_{0\sigma }^{\dagger
}\rangle \rangle _{\omega }$ using equations of motion. Choosing $\varphi
_{j}=0$ for $0<j<L-1$ and $\varphi _{0}=\varphi _{L-1}=\pi /2$, we obtain,
after some algebra for $0\neq l\neq L-1$:

\begin{eqnarray}
j_{l}(\Phi ) &=&\frac{16e}{\pi \hbar L^{2}}tt_{R}t_{L}\sin \Phi
\sum_{n}(-1)^{n}\sum_{n^{\prime }}^{\prime }p(l,n)p(l+1,n^{\prime }) 
\nonumber \\
&&\times \sum_{\sigma }\int d\omega f(\omega )%
\mathop{\rm Im}%
\frac{G_{d\sigma }(\omega +i\eta )}{(\omega +i\eta -\varepsilon _{n})(\omega
+i\eta -\varepsilon _{n^{\prime }})}\text{;}  \nonumber \\
p(l,n) &=&\sin \frac{\pi nl}{L}\sin \frac{\pi n}{L}\text{; }\varepsilon
_{n}=-2t\cos \frac{\pi n}{L}-i\delta .  \label{je}
\end{eqnarray}
Here $\sum_{n}$ runs over all integers $1\leq n\leq L-1$, while $%
\sum_{n^{\prime }}^{\prime }$ is restricted to integers of opposite parity
to that of $n$, $f(\omega )$ is the Fermi function, $\eta $ is a positive
infinitesimal, and $\varepsilon _{n}$ are the eigenergies in absence of the
QD, allowing for a finite broadening $\delta $ \cite{mir}. Eq. (\ref{je}) is
exact. However, we use an approximate $G_{d\sigma }$. Within the precision
of the numerical integration, we have verified that the resulting $j_{l}$ is
independent of $l$, as it should be because of conservation of the current.
Therefore we drop the subscript in what follows.

We calculate $G_{d\sigma }$ using a self-consistent IPA based on
perturbation theory in $U$ up to second order \cite{levy,kaj}, generalized
to allow spin dependence:

\[
G_{d\sigma }^{-1}(\omega )=\left[ G_{d\sigma }^{0}(\omega )\right] ^{-1}-Un_{%
\overline{\sigma }}-\Sigma _{\sigma }(\omega ), 
\]
where $G_{d\sigma }^{0}$ is the Green function for $U=0$ and $E_{d}$
replaced by an effective energy $\varepsilon _{eff}^{\sigma }$ determined
selfconsistently

\begin{eqnarray*}
\left[ G_{d\sigma }^{0}(\omega )\right] ^{-1} &=&\omega -\varepsilon
_{eff}^{\sigma }-\sum_{n}\frac{|V_{n}|^{2}}{\omega -\varepsilon _{n}}\text{; 
} \\
V_{n} &=&\sqrt{\frac{2}{L}}\sin \frac{\pi n}{L}[t_{R}e^{i\Phi
/2}-(-1)^{n}t_{R}e^{-i\Phi /2}],
\end{eqnarray*}
$n_{\sigma }=\langle n_{d\sigma }\rangle $, and:

\[
\Sigma _{\sigma }(\omega )=\frac{n_{\overline{\sigma }}(1-n_{\overline{%
\sigma }})\Sigma _{\sigma }^{(2)}}{n_{\overline{\sigma }}^{0}(1-n_{\overline{%
\sigma }}^{0})-[(1-n_{\overline{\sigma }})U+E_{d}-\varepsilon _{eff}^{\sigma
}]U^{-2}\Sigma _{\sigma }^{(2)}}, 
\]
where $n_{\sigma }^{0}$ is the expectation value of $n_{d\sigma }$
calculated with $G_{d\sigma }^{0}$ and $\Sigma _{\sigma }^{(2)}$ is the
ordinary second order correction to the self energy, calculated from a
Feynmann diagram involving the analytical extension of $G_{d\sigma
}^{0}(\omega )$ to Matsubara frequencies \cite{mir}: 
\begin{eqnarray*}
\Sigma _{\sigma }^{(2)}(i\omega _{l},T) &=&U^{2}T\sum_{m}G_{d\sigma
}^{0}(i\omega _{l}-i\nu _{m})\chi (i\nu _{m})\text{;} \\
\chi (i\nu _{m}) &=&-T\sum_{n}G_{d\bar{\sigma}}^{0}(i\omega _{n})G_{d\bar{%
\sigma}}^{0}(i\omega _{n}+i\nu _{m}).
\end{eqnarray*}
The resulting $G_{d\sigma }$ is not only valid up to $U^{2}$, but it is also
exact for a decoupled dot ($t_{L}=t_{R}=0$), and reproduces the leading term
for $\omega \rightarrow \infty $ \cite{kaj}. We determine $\varepsilon
_{eff}^{\sigma }$ by imposing that $n_{\sigma }^{0}=n_{\sigma }$ for both
spins.

\medskip
\medskip
\medskip
\medskip

\begin{figure}
\narrowtext
\epsfxsize=3.5 truein
\vbox{\hskip 0.05truein \epsffile{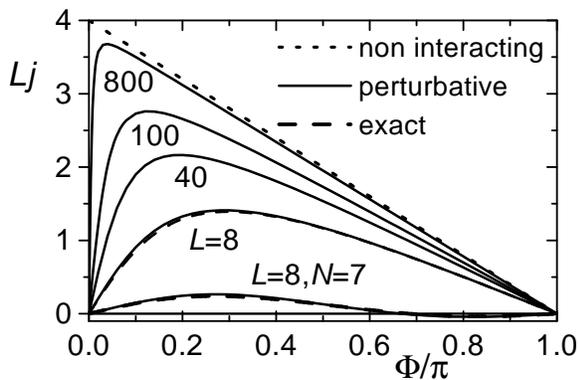}}
\medskip
\caption{Current in units of $et/\hbar $ as a function of magnetic flux for $%
U=2$, $t_{L}=0.4$ and different values of $L$. Unless otherwise indicated $%
N=L$.}
\end{figure}

We take $t=1$ as the unit of energy and keep $E_{d}=-U/2$ and $t_{L}=t_{R}$.
This allows us to exploit electron-hole and reflection symmetry in some
cases. As a basis for our study we choose $U=2$ and $t_{L}=0.4$. For $%
L\rightarrow \infty $ and $\varepsilon _{F}=0$, this leads to a resonant
level width $\Delta =0.32$ \cite{pro} (neglecting its energy dependence).
The ratio $U/\Delta =6.25$ is large enough for the system to be in the Kondo
regime of the model, but low enough to ensure the validity of the IPA \cite
{pro}. The impurity spectral density $\rho _{d\sigma }(\omega )$ shows three
peaks at $E_{d}$, $E_{d}+U$ and $\varepsilon _{F}$ characteristic of the
Kondo regime. For $\delta =0$ and finite $L$, $\rho _{d\sigma }(\omega )$
consists in a set of delta functions. Using \cite{hal}:

\begin{equation}
T_{K}=\sqrt{U\Delta /2}e^{\pi (E_{d}-\varepsilon _{F})(E_{d}-\varepsilon
_{F}+U)/2U\Delta },  \label{tk}
\end{equation}
the above parameters lead to $T_{K}\sim 0.05$ and $\xi \sim 40$. We begin
showing the results for $T=\delta =0$. They depend drastically on the parity
of the number of particles $N$. For even $N$, the results for odd $N/2$ are
essentially the same as those for $N\pm 2$ shifting $\Phi $ by $\pi $. Then
we restrict to either odd $N$ or $N/4$ integer. Using reflection symmetry
around the QD ($c_{j\sigma }\rightarrow c_{L-j\sigma }$), one realizes that $%
j(\Phi )=-j(-\Phi )$. Thus, it is sufficient to represent $j$ in the
interval $0\leq \Phi \leq \pi $. Fig. 1 displays the evolution of $j(\Phi )$
as a function of ring size for even $N$. For $L=8$ and $N=7,8$ we also
compare the IPA results with those obtained using $j(\Phi )=-(eL/\hbar
)\partial E(\Phi )/\partial \Phi $, with the ground state energy $E$
calculated by ED. The maximum deviation between both results takes place
around $\Phi =0.2$ and is below $0.05et/\hbar $. For $L\rightarrow \infty $
and $N$ even, $j(\Phi )$ converges to the non-interacting ideal result, as
predicted by an analysis of the strong-coupling fixed point of RG \cite{aff}%
. This supports the validity of the IPA results for all $L$. Our results
also agree qualitatively with those of RG for $L\ll \xi $, but disagree with
those of Ref. \cite{kang}.

\medskip
\medskip
\medskip
\medskip

\begin{figure}
\narrowtext
\epsfxsize=3.5truein
\vbox{\hskip 0.05truein \epsffile{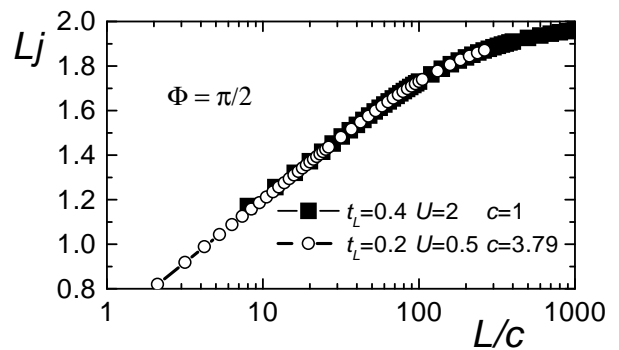}}
\medskip
\caption{Current as a function of $L$ for two sets of parameters.}
\end{figure}

To study the scaling properties in the dependence with $L$, we have
calculated $j(\pi /2)$ for $8\leq L\leq 1000$ and two sets of parameters: $%
U=2$, $t_{L}=0.4$ as before, and $U=0.5$, $t_{L}=0.2$. The new choice should
reduce $\Delta $ by a factor $\sim 4$ \cite{pro}, keeping then nearly the
same value of $U/\Delta $, and resulting in $\sim 4$ times smaller $T_{K}$
(see Eq. (\ref{tk})), and $\sim 4$ times larger $\xi $. In fact, as shown in
Fig. 2, scaling $L$ by a factor 3.79, both functions $Lj(\pi /2,L)$
practically coincide, confirming that $Lj$ is a universal function of $L/\xi 
$, as expected for $L\gg 1$ in the Kondo regime \cite{aff}. For $L\sim 8$
and $U=2$, $Lj$ is shifted a little bit upwards from the universal curve.
For $U=0.5$, using $\xi =151.6$ as derived from Eq. (\ref{tk}), the results
for $8\leq L\leq 240$ fit with negligible errors on the curve:

\begin{equation}
Lj(\pi /2)\hbar /te=1.53+0.24\ln (L/\xi ).  \label{sc}
\end{equation}

\medskip
\medskip
\medskip
\medskip

\begin{figure}
\narrowtext
\epsfxsize=3.5truein
\vbox{\hskip 0.05truein \epsffile{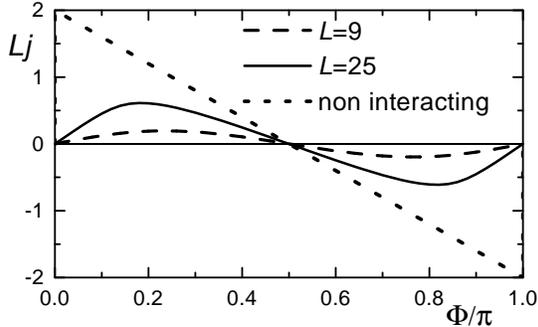}}
\medskip
\caption{Same as Fig. 1 for odd $N$ and $U=1.5$.}
\end{figure}

Using ED, we find that in general for given $\varepsilon _{F}$, the states
with odd $N$ are less stable than those of even $N$. In particular, it is
not possible to find a unique value of $\varepsilon _{F}$ for which some odd 
$N$ are stable for all $\Phi $. Even $N$ with odd (even) $N/2$ are favored
near $\Phi =0$ ($\Phi =\pi $), for which $j(\Phi )$ has a small amplitude.
For example, for $L=12$ and $\varepsilon _{F}=-0.7$, $N$ decreases from 10
for $\Phi =0$ to 8 for $\Phi =\pi $, jumping from 10 to 9 near $\Phi
=0.16\pi $, and from 9 to 8 near $\Phi =0.82\pi $. As a consequence $j(\Phi )
$ is discontinuous and small in magnitude ($|j(\Phi )|<0.5et/\hbar $). The
IPA for odd $N$ was applied imposing a given $N$ (although it might
correspond to a metastable state) and adjusting $\varepsilon _{F}$ as a
function of $\Phi $. A technical difficulty is that for large $U$ and $L$ we
could not find the selfconsistent solution. This is related to the fact that
for finite $L$, $n_{\sigma }^{0}$ and $n_{\sigma }$ are discontinuous
functions of the $\varepsilon _{eff}^{\sigma }$ and it is not always
possible to satisfy $n_{\sigma }^{0}=n_{\sigma }$. As seen in Fig. 1, and
the above mentioned case for $L=12$, $j(\Phi )$ is strongly suppressed for
odd $N$ and small $L$. This is related to a partial suppression of the Kondo
effect. For $L=8$, the expectation value of the impurity spin $%
s_{z}=(n_{\uparrow }-n_{\downarrow })/2$ $\sim 0.4$, indicating only a small
partial screening. Also, $j(\Phi )$ displays positive and negative values in
the interval $0\leq \Phi \leq \pi $, suggesting a tendency to periodicity in 
$\pi $ instead of $2\pi .$ This periodicity is exact for $N=L$ odd, $%
E_{d}=-U/2$ and $t_{L}=t_{R}$: the electron-hole transformation $c_{j\sigma
}^{\dagger }\rightarrow c_{j\sigma }$, $\varphi _{j}\rightarrow \varphi
_{j}+\pi $, maps $H(-\Phi )$ onto $H(\Phi +\pi )$ and $j(-\Phi )$ onto $%
-j(\Phi +\pi )$ (see Eqs. (\ref{he}) and (\ref{j})). Combining this with $%
j(\Phi )=-j(-\Phi )$, one has $j(\Phi )=j(\Phi +\pi )$, and $j(\pi /2)=0$.
In Fig. 3 we show $j(\Phi )$ in two of these cases with $U$ reduced to 1.5
to be able to obtain self consistent solutions up to $L=37$. We estimate $%
\xi \sim 25$. According to RG arguments, as $L$ increases, $j(\Phi )$ tends
to the result of a fictitious non-interacting system with spin dependent $%
\varepsilon _{F}$ to allow for odd $N.$ Our results are consistent with
this. $s_{z}$ decreases from $\sim 0.37$ to $\sim 0.15$ as $L$ increase from
9 to 37.

\medskip
\medskip
\medskip
\medskip

\begin{figure}
\narrowtext
\epsfxsize=3.5truein
\vbox{\hskip 0.05truein \epsffile{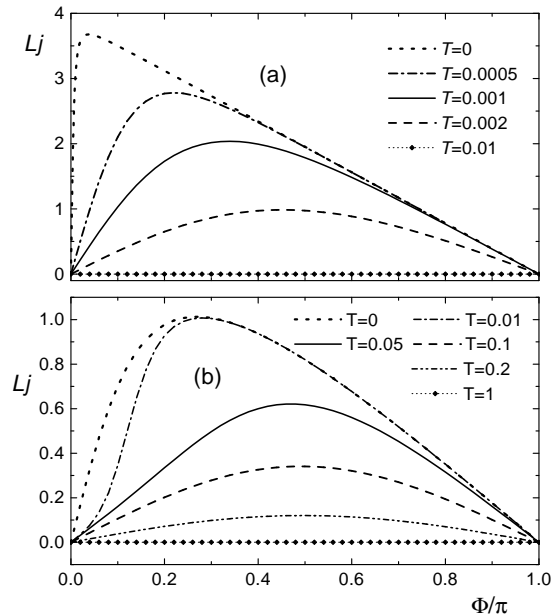}}
\medskip
\caption{Same as Fig. 1 for different temperatures and (a) $U=2$, $t_{L}=0.4$%
, $L=800$, (b) $U=0.5$, $t_{L}=0.2$, $L=8$.}
\end{figure}

We have calculated the $T$ dependence of $j(\Phi )$ for even $N$ and several 
$L/\xi $. In Fig. 4, we show  two cases: (a) $T_{K}\sim 0.05$, $L/\xi \sim 20
$ and (b) $T_{K}\sim 0.012$, $L/\xi \sim 1/20$. From the known results for
the conductance through a QD \cite{wiel,pro} one would expect $T_{K}$ to be
the relevant scale for the $T$ dependence. However, it is a fraction of the
level spacing $D=2\pi \hbar v_{F}/L$ in both cases. This is easy to
understand in case (a): for $T\ll T_{K}$ and $L\gg \xi $, the physics is
still dominated by the strong coupling fixed point of RG for which the model
reduces to a non-interacting one \cite{aff}. In turn, the conductance of the
latter is strongly reduced for $T\sim D\ll T_{K}$. In contrast to the
infinite system, for which the Kondo peak in $\rho _{d\sigma }(\omega )$
looses approximately half its intensity for $T\sim T_{K}$ \cite{pro}, when $%
T_{K}\ll D$ ($L\ll \xi $), $G_{d\sigma }$ is practically not modified with
increasing $T$, until $T$ reaches a sizeable fraction of $D$. As a
consequence for all $L$ the scale for the $T$ dependence is $\sim D/5$, but
depends on $\Phi $. Note that these arguments are independent of the parity
of $N$. The last argument is in agreement with recent RG calculations, which
show that for even $N$ and $T_{K}\ll D$, the screening of the localized spin
takes place at $T\sim D$, while as discussed above, the screening is only
partial for odd $N$ \cite{cor}. In other words decreasing $L$, the Kondo
effect is enhanced for $N$ even and inhibited for $N$ odd.

\medskip
\medskip
\medskip
\medskip

\begin{figure}
\narrowtext
\epsfxsize=3.5truein
\vbox{\hskip 0.05truein \epsffile{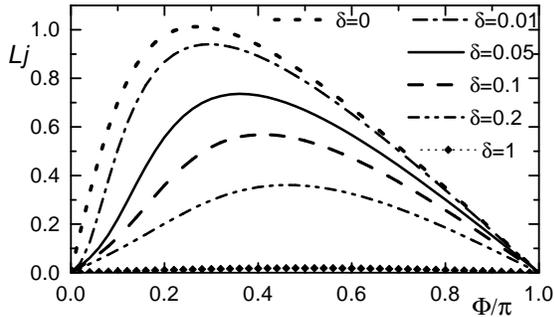}}
\medskip
\caption{Same as Fig. 4 (b) for $T=0$ and several values of $\delta $.}
\end{figure}

So far, we have considered $\delta =0$ and $G_{d\sigma }(\omega )$. However,
real mesoscopic systems are coupled to a charge reservoir which determines $%
\varepsilon _{F}$ and discrete electronic states become resonances. The
effects of this coupling have been studied recently \cite{mir,cor}. The
coupling to the conduction eigenstates (or to the impurity) is essential to
explain the line shape in the projection of the Kondo effect to a remote
location (mirage effect) \cite{mir}. In the case $D\ll T_{K}$, it is clear
that $j$ is strongly reduced when $\delta $ reaches $D$. In the opposite
case $T_{K}\ll D$, for $N$ even (corresponding to some $\varepsilon _{n}$
near $\varepsilon _{F}$), $\rho _{d\sigma }(\omega )$ near $\varepsilon _{F}$
evolves with increasing $\delta $ from two delta functions at both sides of $%
\varepsilon _{F}$ to one peak centered at $\varepsilon _{F}$ \cite{mir}. The
corresponding change in $j(\Phi )$ is shown in Fig. 5. The effect of the
broadening of the levels is similar to that of increasing $T$, but a little
bit weaker and more evenly distributed in $\Phi $.

We discuss briefly the side dot. The Hamiltonian is:

\begin{eqnarray}
H &=&-\sum_{\sigma ,j=0}^{L-1}\left( te^{i\varphi }c_{j+1\sigma }^{\dagger
}c_{j\sigma }+\text{H.c.}\right) -t_{d}\sum_{\sigma }(d_{\sigma }^{\dagger
}c_{0\sigma }+\text{H.c.})  \nonumber \\
&&+E_{d}\sum_{\sigma }n_{d\sigma }+Un_{d\uparrow }n_{d\downarrow },
\label{hs}
\end{eqnarray}
where now $\varphi =\Phi /L$ and $n_{d\sigma }=d_{\sigma }^{\dagger
}d_{\sigma }$. Proceeding as before, for $\delta =0$ the current becomes:

\begin{eqnarray}
j_{s}(\Phi ) &=&\frac{2te}{\pi \hbar L^{2}}\{-2\pi L\sum_{k}\sin (k+\varphi
)f(\varepsilon _{k})  \nonumber \\
&&+t_{d}^{2}\sum_{k}\sum_{k^{\prime }}\sin [\varphi +k^{\prime }(l+1)-kl] \nonumber \\
&&\times \sum_{\sigma }\int d\omega f(\omega )  
\mathop{\rm Im}%
\frac{G_{d\sigma }(\omega +i\eta )}{(\omega +i\eta -\varepsilon _{k})(\omega
+i\eta -\varepsilon _{k^{\prime }})}\}\text{;}  \nonumber \\
\varepsilon _{k} &=&-2t\cos (k+\varphi )\text{,}  \label{js}
\end{eqnarray}
independently of $l$. The expressions for $G_{d\sigma }$ are the same as
before with $V_{n}$ replaced by $V_{k}=1/\sqrt{L}$. In the limit $%
t_{d}\rightarrow 0$, $L\rightarrow \infty $, the resulting resonant level
width $\Delta $ is the same as that of an embedded dot with $%
t_{L}=t_{R}=t_{d}/2$ \cite{pro}. Then, we take $t_{d}=0.8$, $N=L+1$ even and
other parameters as in Fig. 1 to study the dependence of $j_{s}(\Phi )$ on $L
$. The result is shown in Fig. 6. In agreement with Ref. \cite{aff} $%
j_{s}(\Phi )$ is suppressed for large $L$. Note that the first term of Eq. (%
\ref{js}) is the non-interacting one, which is of the order of 100 times $%
j_{s}(\Phi )$ for $L=799$. The fact that the small $j_{s}(\Phi )$ comes from
a near cancellation of two terms, and only the second one is approximate,
further supports the validity of the IPA.

\medskip
\medskip
\medskip
\medskip

\begin{figure}
\narrowtext
\epsfxsize=3.5truein
\vbox{\hskip 0.05truein \epsffile{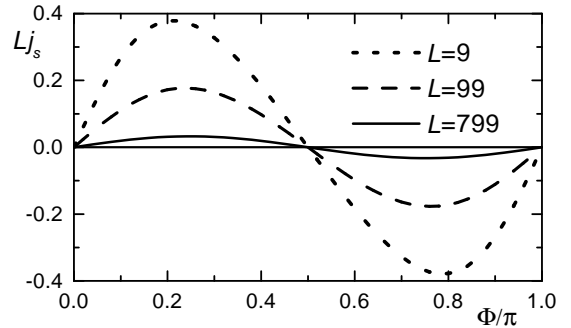}}
\medskip
\caption{Current for the side dot as a function of flux for $U=2$, $%
t_{d}=0.8 $, $T=\delta =0$, $N=L+1$ and several values of $L$.}
\end{figure}

We have calculated the persistent current $j(\Phi )$ in a ring with a QD
using an interpolative perturbative approach. The method is accurate enough
to describe $j(\Phi )$ for all $L/\xi $.  The
universal dependence of $Lj$ with $L/\xi $ is displayed. The energy scale
for the dependence on temperature and broadening of the levels is a fraction
of the level spacing $D$ instead of $T_{K}$. For the side dot $j_{s}(\Phi )$
is small and decreases with $L$. For both systems and odd $L$, the period of 
$j(\Phi )$ is $\pi $ in half filled symmetric rings. Because of their larger
stability and larger currents, states with even number of electrons $N$ in
embedded dots seem the most interesting experimentally.

We thank G. Chiappe and C.R. Proetto for useful discussions. We are
partially supported by CONICET. This work was sponsored by PICT 03-03833
from ANPCyT.


\begin{references}
\bibitem{wiel}  W.G. van der Wiel {\it et al.}, Science {\bf 289}, 2105
(2000).

\bibitem{schu}  R. Schuster {\it et al.}, Nature (London) {\bf 385}, 417
(1997).

\bibitem{hei}  Y. Yi M. Heiblum {\it et al.}, Science {\bf 290}, 779 (2000).

\bibitem{zv1} A.A. Zvyagin and T.V. Bandos, Low Temp. Phys. {\bf 20}, 222
(1994); A.A. Zvyagin, {\it ibid} {\bf 21}, 349 (1995).

\bibitem{fer}  V. Ferrari {\it et al.}, Phys. Rev. Lett. {\bf 82}, 5088
(1999).

\bibitem{kang}  K. Kang and S. C. Shin, Phys. Rev. Lett. {\bf 85}, 5619
(2000).

\bibitem{aff}  I. Affleck and P. Simon, Phys. Rev. Lett. {\bf 86}, 2854
(2001).

\bibitem{hu}  H. Hu, G-M Zhang, and L. Yu, {\bf 86}, 5558 (2001).


\bibitem{eck}  H.-P. Eckle, H. Johannesson, and C.A. Stafford, Phys. Rev.
Lett. {\bf 87}, 016602 (2001).

\bibitem{anda}  E.V. Anda {\it et al.}, cond-mat/0106055.

\bibitem{zv2} A.A. Zvyagin, , cond-mat/0203253.

\bibitem{mir}  A.A. Aligia, Phys. Rev. B {\bf 64}, 121102(R) (2001).

\bibitem{levy}  A. Levy-Yeyati, A. Mart\'{\i }n-Rodero, and F. Flores, Phys.
Rev. Lett. {\bf 71}, 2991 (1993); references therein.

\bibitem{kaj}  H. Kajueter and G. Kotliar, Phys. Rev. Lett. {\bf 77}, 131
(1996).

\bibitem{pro}  A.A. Aligia and C.R. Proetto, Phys. Rev. B {\bf 65}, 165305
(2002).

\bibitem{hal}  F.D.M. Haldane, Phys. Rev. Lett. {\bf 40}, 416 (1978).

\bibitem{cor}  P.S. Cornaglia and C.A. Balseiro, cond-mat/0202489.
\end{references}
\end{document}